\documentclass[%
reprint,
superscriptaddress,
amsmath,amssymb,
aps,
pra,
er,
floatfix,
]{revtex4-1}

\usepackage{graphicx}
\usepackage{dcolumn}
\usepackage{bm}
\usepackage[hidelinks]{hyperref}

\usepackage{physics}
\usepackage{siunitx}
\usepackage[version=4]{mhchem}
\usepackage{color}
\usepackage{comment}
\DeclareSIUnit\angstrom{\mbox{\normalfont\AA}}
\DeclareSIUnit\bar{bar}

\begin{document}

\preprint{APS/123-QED}

\author{Riccardo Foffi}
\affiliation{Institute for Environmental Engineering, Department of Civil, Environmental and Geomatic Engineering, ETH Z\"urich, Laura-Hezner-Weg 7, 8093 Z\"urich, Switzerland.}
\author{Francesco Sciortino}
\email[To whom correspondence should be addressed. Email: ]{francesco.sciortino@uniroma1.it}
\affiliation{Dipartimento di Fisica, Sapienza Universit\`a di Roma, Piazzale Aldo Moro 5, I-00185 Rome, Italy.}

\title{Identification of local structures in water from supercooled to ambient conditions}


\begin{abstract}
Studies of water thermodynamics have long been tied to the identification of two distinct families of local structures, whose competition could explain the origin of the many thermodynamic anomalies and of the hypothesized liquid-liquid critical point in water.
Despite the many successes and insights gained, the structural indicators proposed throughout the years were not able to unequivocally identify these two families over a wide range of conditions.
We show that a recently introduced indicator, $\Psi$, which exploits information on the HB network connectivity, can reliably identify these two distinct local environments over a wide range of thermodynamic conditions (188 to 300 K and 0 to 13 kbar), and that close to the liquid-liquid critical point the spatial correlations of density fluctuations are identical to those of the $\Psi$ indicator.
Our results strongly support the idea that water thermodynamic properties arise from the competition between two distinct and identifiable local environments.
\end{abstract}

\maketitle

\section{Introduction}

Water is a liquid with fascinating physical properties~\cite{debenedetti2003supercooled,gallo2016water,handle2017supercooled,tanaka2020liquid}.  Differently from many other common compounds, the  thermodynamics response functions  of liquid water display a non-monotonic  temperature and pressure dependence. For example, at ambient pressure,  the compressibility has a minimum at \SI{46}{\celsius} and a maximum at approximately $\SI{-43}{\celsius}$, and the density has a maximum at \SI{4}{\celsius}~\cite{gallo2016water,kim2017maxima}. 
These non-monotonic behaviors  are strongly suggestive of relevant structural changes taking place in the liquid state. 
  
 The peculiarities of water can be traced back to the  strength and  directionality of the hydrogen-bond interaction and to the limited number of hydrogen-bonds that a water molecule can form with its neighbours~\cite{russo2022physics,smallenburg2015tuning}. Unlike other substances, water molecules can assume a variety of local structures, from  the highly tetrahedral open
 configuration in which the water molecule participates in four linear hydrogen bonds, to 
more distorted and denser local environments.  The radial distribution function of the oxygen atoms indeed reveals the presence of 
interstitial molecules, located between the first and the second tetrahedral shells~\cite{soper2000structures,skinner2016structure}. 
 There is consensus that
molecules in ``tetrahedral" configurations are characterized by low energy,
local order, low local density, while molecules at the other extreme are 
characterized by higher energies, higher disorder, higher local density~\cite{tanaka2000simple,holten2012entropydriven,biddle2017twostructure}. 
If this large variety of configurations can be grouped in two families or if it reflects a
continuum of geometries is still object of controversy~\cite{niskanen2019compatibility,soper2019water,johari2015thermodynamic}.

The idea of two families of different local environments is consistent with the hypothized presence of a liquid-liquid critical point~\cite{poole1992phase} in supercooled states, the end point of a line of liquid-liquid first order transitions.
Such a critical point, originally discovered in the ST2 water model~\cite{poole1992phase}, has recently been confirmed with accurate free-energy calculations in several high-quality classical potentials (TIP4P/2005, TIP4P/Ice~\cite{debenedetti2020second}, WAIL~\cite{weis2022liquid}) 
significantly strengthening the possibility that such unconventional thermodynamic scenario is  representative  of real water.  Neural network potentials based on quantum mechanics calculations~\cite{gartner2020signatures} and path-integral simulations~\cite{eltareb2022evidence} also support the presence of such liquid-liquid critical point. 
Several experiments also support the liquid-liquid critical point scenario~\cite{mishima1998decompressioninduced,winkel2011equilibrated,kim2017maxima}.
In particular, recent X-rays scattering experiments probing sub-microsecond timescales to observe the
relaxation of the metastable liquid before nucleation have   provided evidence of a transition  between two different structures~\cite{kim2020experimental,amann-winkel2023liquidliquid} in deep supercooled states.

The presence of a  critical point  requires the competition between two  different local structures  and  a free-energy gain when local structures of the same type cluster in space.  Several studies have shown that
a simple two-state description of the free-energy~\cite{cuthbertson2011mixturelike, tanaka2000simple,
holten2012entropydriven, li2013liquid,
singh2016twostate,
biddle2017twostructure,
caupin2021minimal,yu2023unified}, in which the entropic term of
mixing two different local structures (differing in energy, entropy and density) 
is complemented by a clustering contribution is able to describe the equation of states of 
several numerically studied models, as well as reproduce the equation of states  of water better than any other previously proposed expression~\cite{anisimov2018thermodynamics}.

Numerical simulations have been thoroughly  scrutinised searching for  quantities that could detect the two families of molecular structures~\cite{ 
errington2001relationship,
cuthbertson2011mixturelike,
shiratani1996growth,
russo2014understanding,
tanaka2019revealing,
montesdeoca2020structural, verde2021structural,
muthachikavil2022structural}. Unfortunately, most proposed  indicators (but see for exceptions Refs.~\cite{cuthbertson2011mixturelike,montesdeoca2020structural, verde2021structural}), while based on strong physical intuitions,  have typically resulted into wide distributions --- which in some cases can be represented by the superposition of two distributions with  relative weights changing  with pressure and/or temperature. A
 clear indication of a well separated two-state behavior  (a distribution function with two well resolved peaks) has remained elusive.  The ongoing attempt to include longer-range informations in the definition of the structural indicators~\cite{faccio2022low,neophytou2022topological,loubet2023turning,daidone2023statistical}, despite its sound physical expectation, has not drastically improved the classification. 

In the attempt to better characterize the features of the radial distribution function $g(r)$, we have recently developed~\cite{foffi2021structural,foffi2021structure} an analysis connecting the physical distance between two
molecules with their chemical distance, measured as the (smallest) number of hydrogen bonds  that need to be crossed on going from one molecule to the other. All structural features appearing in $g(r)$ were thus associated to specific bonding geometries. In particular, the
interstitial molecules, the ones populating the region around 3.5 $\si{\AA}$ between the first and second peak of the $g(r)$,  have been 
associated to molecules with chemical distance larger or equal to four. We have also found that a molecular order parameter (that we named $\Psi_i$), defined for each molecule $i$ as
 the smallest distance in real space among all molecules with chemical distance 
four has, close to the critical point, a well defined bimodal distribution function.
Tetrahedral, low-density local structures were found to be  characterised by large values of $\Psi_i$ ($ \sim 6.5 \si{\AA} $), while high-density local structures favour shorter distances ($ \sim{3.5}\si{\AA}  $). Even more, at  pressures   close to the critical point, the average over all molecules in the system $\expval{\Psi_i}$ fluctuates  exactly as the density~\cite{foffi2023correlated}, confirming that $\expval{\Psi_i}$ could very well  be chosen as order parameter of the liquid-liquid transition. 

In this article we extend this analysis to a wide range of temperatures (from 180 to 300 K)  and pressures (from 1 to 2500 bar), to quantify the temperature and pressure dependence of this indicator.   At the lowest temperature, we extend the analysis to the very high density region (up to pressures of 13 kbar), presenting an interpretation of the VHDA glass as the limiting structure composed by $\Psi_i \sim{3.5}\si{\AA} $ molecules.

\section{Methods}
Most of the trajectories analyzed in this article have been previously generated~\cite{foffi2021structural,foffi2021structure}  using GROMACS 5.1.4~\cite{abraham2015gromacs} in the NPT ensemble (Nos\'e-Hoover thermostat and Parrinello-Rahman barostat) and reproduce the dynamics of a  system of 1000 rigid water molecules interacting via the TIP4P/Ice model~\cite{abascal2005potential}.   To suppress thermal vibration that are known to blur structural properties and hydrogen-bond identification, we have
calculated the inherent structures (IS)~\cite{sciortino2005potential}, the local potential energy minima,   via a constant volume steepest descent path starting from  equilibrated configurations.
 As previously done~\cite{foffi2021structural,foffi2021structure}, we find the IS using the steepest descent algorithm in GROMACS. The reproducibility of our MD simulations was tested against other TIP4P/Ice results recently reported by \citet{lupi2021dynamical,espinosa2023possible} (see SI).
The presence of an hydrogen bond between two molecules is detected via the Luzar-Chandler geometric criterion~\cite{luzar1993structure}. In short, two molecules are hydrogen-bonded if the H-O-H angle is smaller than
\SI{30}{\degree} and the oxygen-oxygen distance is smaller than 3.5 \si{\AA}.  When applied to IS,
this criterion  can properly identify the hydrogen bonds in the system over an extremely wide range of temperatures and pressures~\cite{foffi2021structure}.  

To evaluate the $\Psi$ structural indicator, the chemical distance $D$ between any pair of molecules $i$ and $j$ is calculated by 
counting the minimum number of hydrogen bonds which needs to be traveled  to move along the HB network  from site  $i$ to site $j$.
As we will show the resulting $\Psi$ histograms can be for convenience represented as a binary mixture of Burr Type XII distributions~\cite{burr1942cumulative}.

We also analyze new simulations of a system of 250000 water molecules in the NVT ensemble at density 1.015 g/cm$^3$ and $T=\SI{191}{\K}$. This state point, on the critical isochore, is quite close to the critical point, such that density fluctuations at small wavevectors can be observed and correlated with the fluctuations of $\expval{\Psi}$.

\begin{figure}[h]
    \centering
	\includegraphics[width=0.4\textwidth]{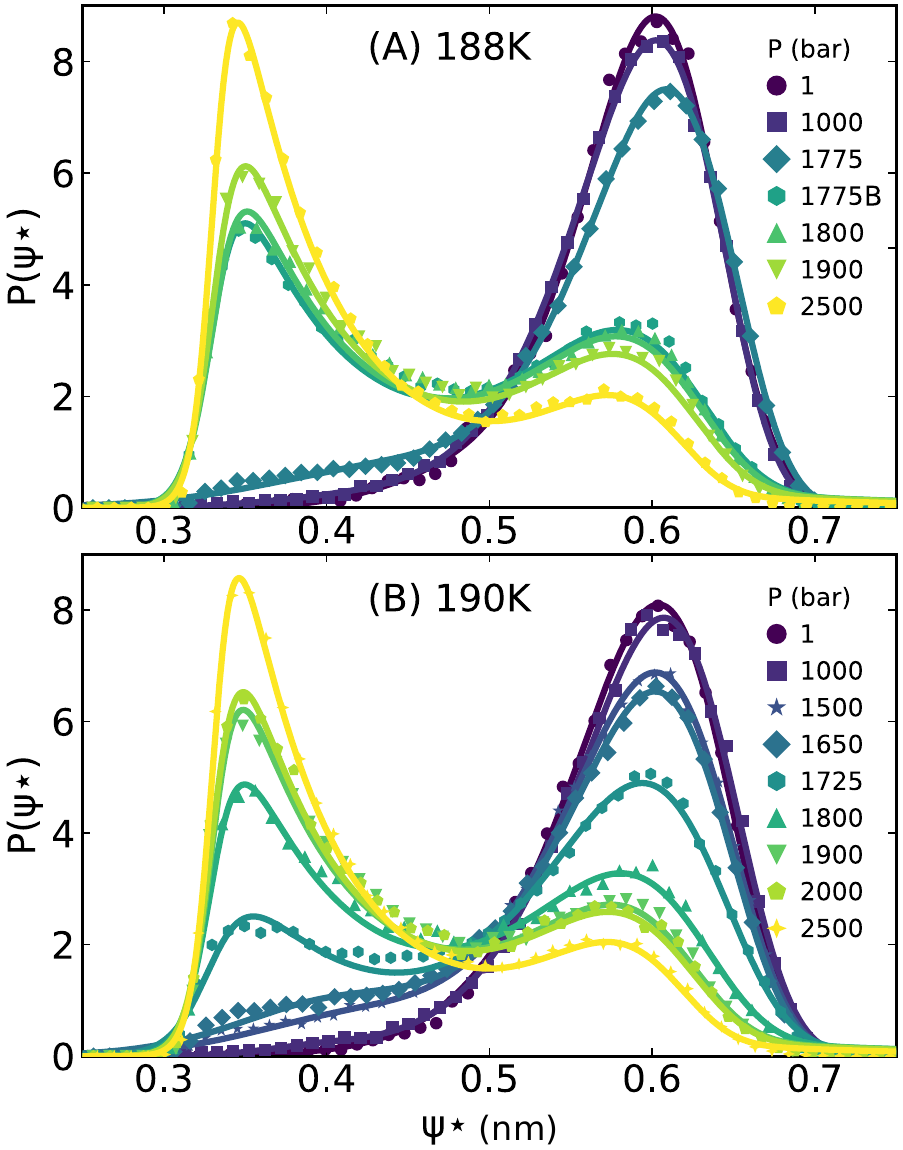}
	\caption{Pressure dependence of the $\Psi^*$ distribution
    along (A) a sub-critical isotherm $T=\SI{188}{K}$ and
    (B) a super-critical isotherm $T=\SI{190}{K}$.
    A clear bimodal behavior is manifested.
    The  simulation labeled 1775B was initialized from a high-density configuration
    as opposed to the one labeled 1775 bar which was initialized from a low-density configuration.
    Points represent simulation data and solid lines the regression from Eq.~\ref{eq:burr}.}
	\label{fig:psiT188T190}
\end{figure}

\begin{figure*}
    \centering
	\includegraphics[width=.9\textwidth]{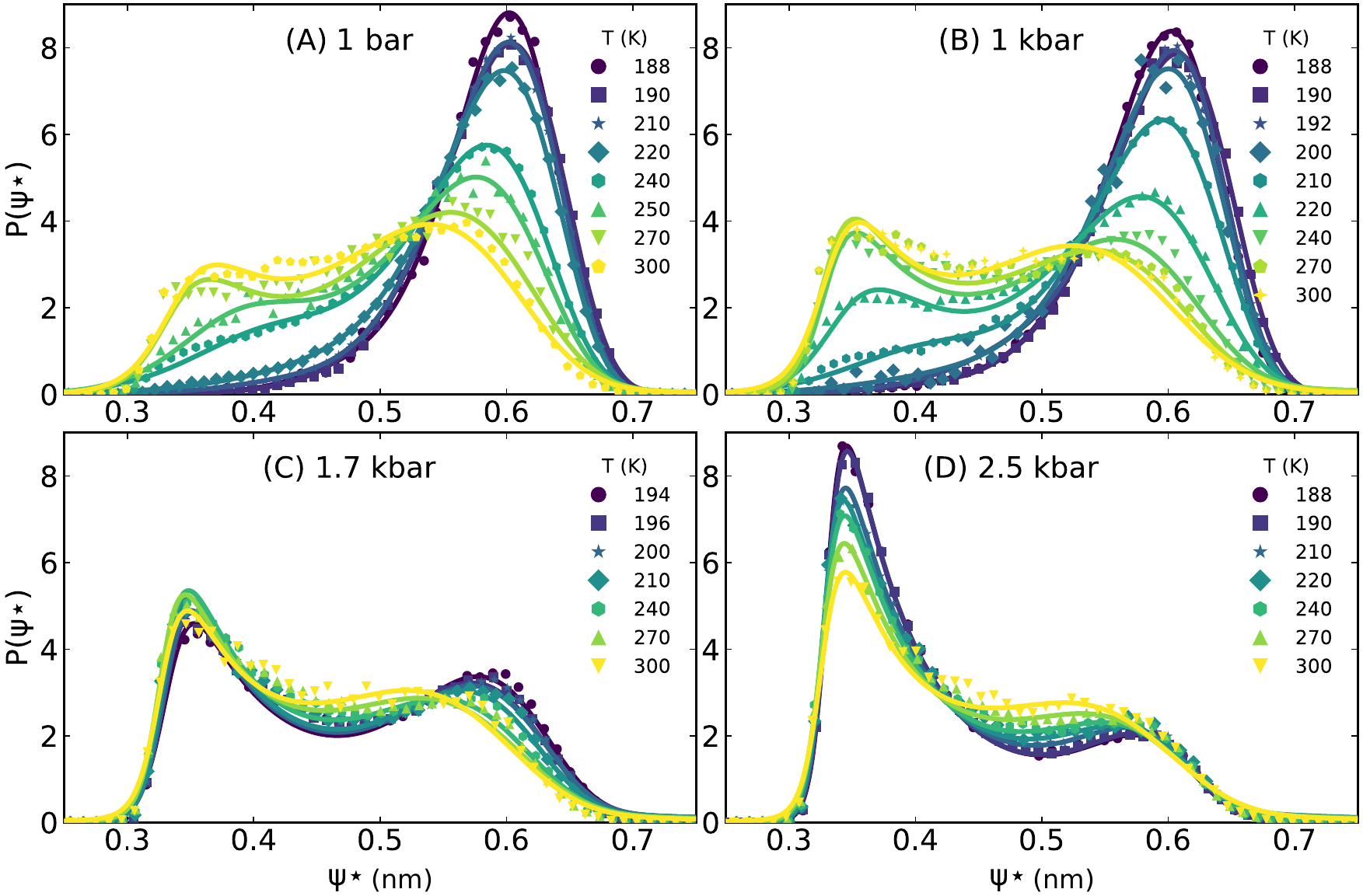}
	\caption{Temperature dependence of $P(\Psi^*)$ for four different pressure values.
    Points represent simulation data and solid lines the regression from Eq.~\ref{eq:burr}.}
	\label{fig:psivarieP}
\end{figure*}

\section{Results and discussion}

It is well established that, as water is exposed to increasing pressures, the Oxygen-Oxygen radial distribution function shows an increasing signal at separations around \SI{3.5}{\angstrom}, a distance which lies between the peaks of the first (\SI{2.8}{\angstrom}) and second(\SI{4.5}{\angstrom}) spatial shells of the liquid at ambient conditions~\cite{soper2000structures,skinner2016structure}.
This feature, which signals a progressive distorsion of the open tetrahedral geometry, has been well characterized over a broad range of conditions and becomes particularly evident in the amorphous ices, when, upon crossing from HDA to VHDA, an initially minor signal in this ``interstitial'' region, transforms into a dominant peak, highlighting a drastic restructuring of the H-bond network~\cite{mariedahl2018xray,foffi2021structure}.
Interestingly, this interstitial population has been shown to arise from molecules which are at a chemical distance of four (or more) from the central one~\cite{foffi2021structural}.  It is important to stress that most of these interstitial molecules are still involved in four hydrogen bonds, albeit more distorted on average than in the open tetrahedral environment, and therefore they cannot be directly associated with coordination defects  (c.f. three or five coordinated molecules) in the HB network, although some correlation exists~\cite{foffi2021structural,loubet2023turning}).  In a previous work~\cite{foffi2021structural}, we have shown that the distribution of real space distances between molecules at chemical distance four displays a clear separation between a
group of molecules located around \SI{3.5}{\angstrom} and anoter group with distances $\approx\SI{8}{\angstrom}$.
This significant separation in real space 
($3.5$ vs. $8.0\,\si{\AA}$) has been exploited to build, in Ref.~\cite{foffi2023correlated},
a molecular indicator expressly designed to quantify the local environment of each molecule.  More precisely,   $\Psi_i$ is defined as the minimum distance in real space between all pairs $i-j$ , where the index $j$ runs over all molecules at chemical distance four from $i$.  Once a proper definition of H-bond is accepted, this new indicator $\Psi_i$ does not require 
any arbitrary cut-off in its definition, eliminating the possibility of cut-off dependent 
findings. In Ref.~\cite{foffi2023correlated} it was shown that averaging 
$\Psi_i$ over all molecules in the system produces a global (as opposed to local) indicator 
which accurately describes the critical fluctuations in the vicinity of the
liquid-liquid critical point, which, for the TIP4P/Ice model, was estimated at $T_c\approx\SI{188.6}{\K}$, $P_c\approx\SI{1750}{\bar}$, $\rho_c\approx\SI{1.015}{\g\per\centi\m^3}$~\cite{debenedetti2020second}.

Fig.~\ref{fig:psiT188T190} shows the distribution of $\Psi_i$ for different pressures
at deep supercooled conditions, below (T=188 K) and above (T=190 K) the critical temperature $T_c  \approx \SI{188.6}{K}$~\cite{debenedetti2020second}.  To subtract the trivial isotropic scaling component in the relative distances on varying the density, we show the distribution as a function of $\Psi^* \equiv \Psi [\frac{\rho(T,P)}{\SI{1}{g/cm^3}}]^{1/3}$, where $\rho(T,P)$ is the  temperature and pressure dependent density.   At low pressures ($P<1000$ bar), the distribution is asymmetric and centred  around $\Psi^* \approx 0.6$, with a  negligible tail  below $\Psi^* \approx 0.4$.  On increasing pressure, this last region starts to be populated.  Below $T_c$ the distribution jumps 
from the low density to the high density liquid value, while the same change is
observed progressively at temperature above (but close to) $T_c$. The cross-over from the open tetrahedral distance  $\Psi^* \approx 0.6$ to the interstitial distance  $\Psi^* \approx 0.35$ is clearly detectable from the distribution functions.  Beside the cross-over, the distributions show a marked two-peak behavior, with an approximate crossing (isosbestic) point  around $\Psi^* \approx 0.5$.   The clear two-peaks structure provides an indisputable evidence of two  major local environments  characterizing supercooled water, significantly reinforcing the underlying idea on which  two-state models have been developed in the past.    Data in Fig.~\ref{fig:psiT188T190} also show that, as expected,
the two coexisting liquids (data for $P=1775$ bar at T=188 K), are not "pure", the low-density liquid containing a small fraction of interstitial molecules and, vice versa,
the high-density liquid containing a non-negligible fraction of tetrahedral local structures.

To highlight the temperature dependence
next we consider the behavior of $\Psi^*$ along four different isobars
 (1,1000,1700, 2500 bar) from 300 K down to $T_c$.  The two lowest isobars are characterized by a significant change in density on cooling.  
Fig.~\ref{fig:psivarieP} shows that the change in density is accompanied by a
significant change in the structure of the liquid, as revealed by the distributions of
$\Psi^*$.   Both for $P=1$ and $P=1000$ bar, the fraction of interstitial molecules
decreases on cooling, almost vanishing at the lowest temperature, consistent with the
expectation that interstitial local configurations are characterized by a higher energy 
per molecule.  The density change with temperature at $P=1700$ and $P=2500$ bar
is significantly more limited and correspondingly, the $\Psi^*$ distributions do not show a significant change.  At all state points the two-peaks structure is very evident, confirming that
the two families of local environments are already very well characterised also
at ambient temperatures. 
We stress that above the critical point there is a single free energy minimum in supercritical conditions, which encompasses two structurally distinct families: at the single molecule level, the two families unequivocally exist as shown by the bimodality of $P(\Psi)$.

\begin{figure}
	\includegraphics[width=0.45\textwidth]{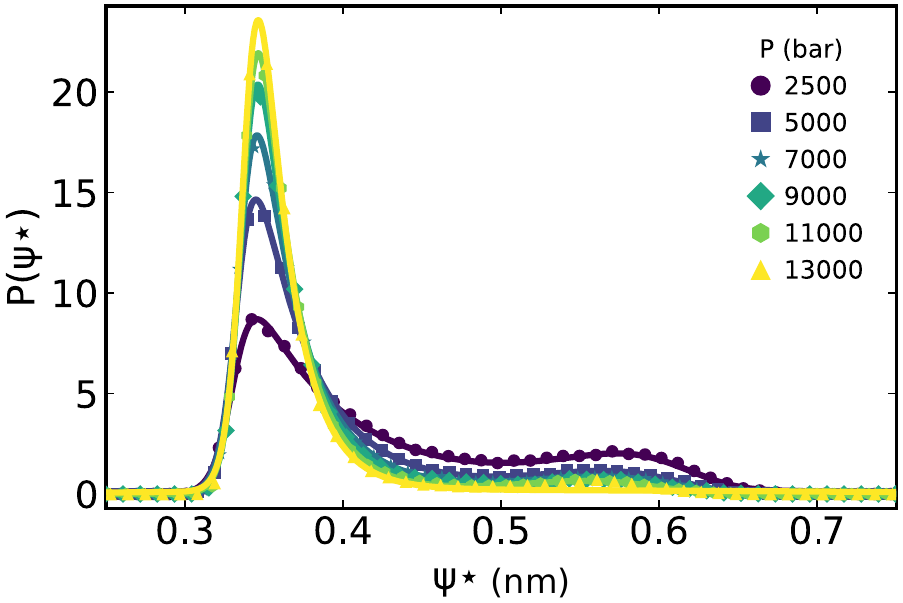}
	\caption{Cross-over from the high-density liquid to the very high-density structure as reflected by the $P(\Psi^*)$  distribution at $T=188$ K.  In a continuous way, all molecules become surrounded by interstitial molecules.
    Points represent simulation data and solid lines the regression from Eq.~\ref{eq:burr}.}
	\label{fig:hdatovhda}
\end{figure}

The data in Fig.~\ref{fig:psivarieP} show that a pure tetrahedral system is
reached at very deep supercooling at ambient pressure. The opposite limit,
in which essentially all molecules have interstitial neighbours is found in the
very-high density limit, again at low temperature. To show this we 
 follow at $T=188$ K the evolution of the distribution of $\Psi$ 
with pressure also in the region between 2000 and 13000 bar, where
the high density liquid continuously transforms into the very high density structure.
This process, described in Fig.~\ref{fig:hdatovhda}, is accompanied by the
progressive disappearance of molecules with large $\Psi$.  Around 13000 bar, 
all molecules have at least one interstitial neighbour and $P(\Psi^*)$ peaks around 0.35.

\begin{figure}
    \centering
    \includegraphics[width=0.45\textwidth]{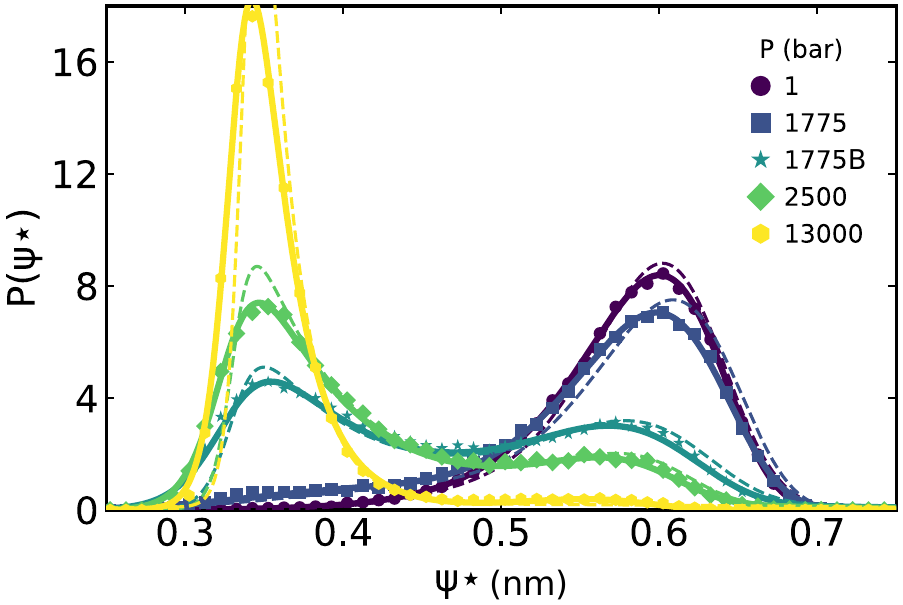}
    \caption{The distribution of $\Psi^*$ evaluated in the real dynamics conserves the bimodality and the wide separation of the two local states that is observed in the inherent structures. Shown here selected points along the $T=\SI{188}{K}$ isotherm.
    The thin dashed lines represent the distributions at the corresponding thermodynamic states in the inherent structures.}
    \label{fig:psird}
\end{figure}

\begin{figure}
  \begin{center}
    \includegraphics[width=0.45\textwidth]{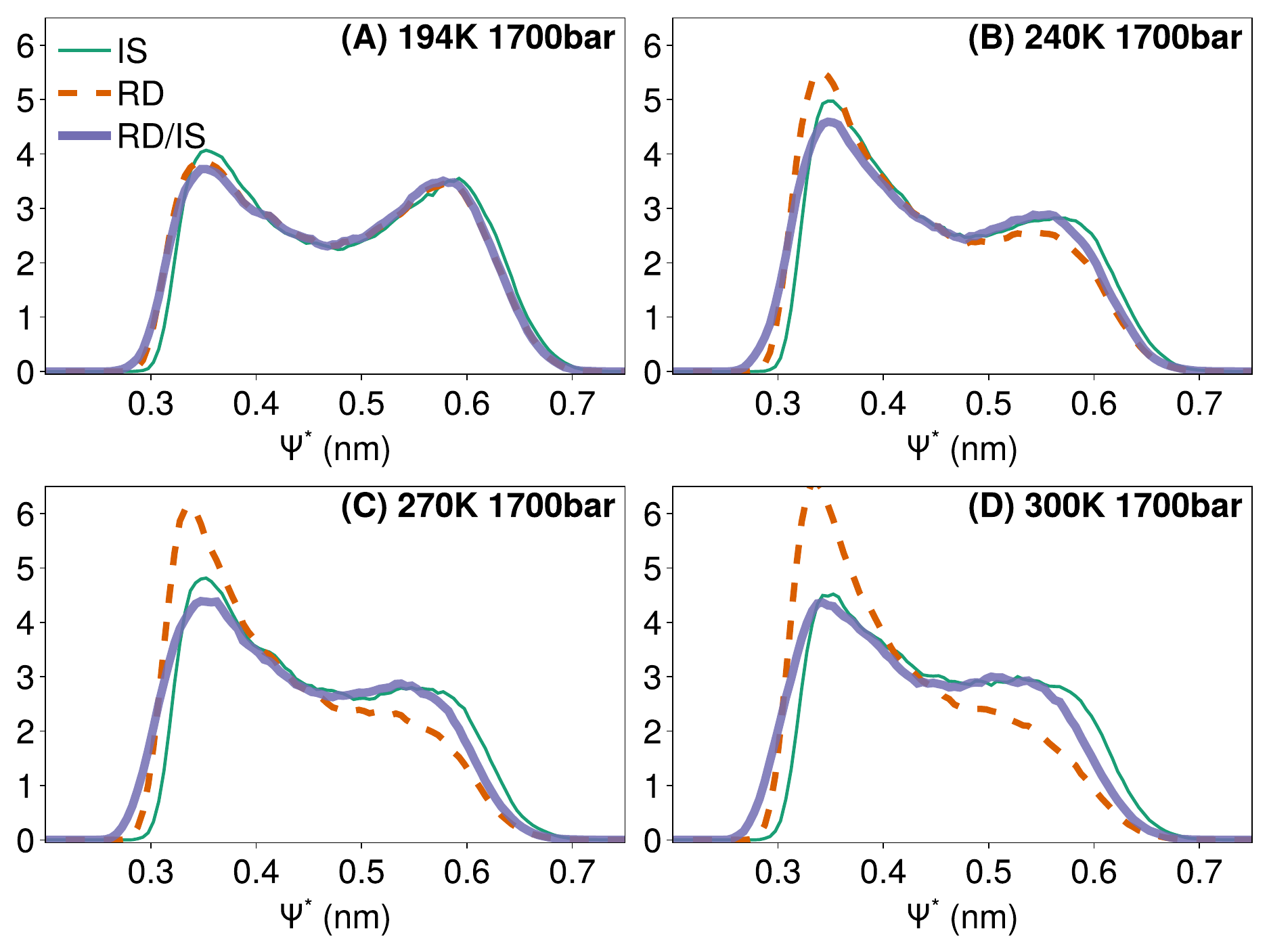}
  \end{center}
  \caption{The distribution of $\Psi^*$ evaluated from \SI{194}{\K} to \SI{300}{\K} along the \SI{1700}{\bar} isobar, shows no significant differences between inherent structures and real dyamics, provided that H-bonds are properly identified.
  Curves labeled ``IS''  and ``RD'' (thin green and dashed orange, respectively), correspond to $\Psi^*$ distributions evaluated in the inherent structures and real dynamics.
  Curves labeled ``RD/IS'', in thick violet, have been evaluated using the spatial configuration of the RD, but employing the definition of the H-bond network obtained from the IS.}
  \label{fig:psird-isobar}
\end{figure}

We also observe that $\Psi$, unlike other indicators, is not critically affected 
by being evaluated using the  inherent structure coordinates. When evaluated in the real dynamics (i.e., the structures directly sampled during the simulation, before the suppression of thermal vibrations via energy minimization), the features of $\Psi$ at low $T$ are fundamentally unchanged (see Fig.~\ref{fig:psird}).
The peaks trivially display a slight broadening due to the vibrational noise (and hence also a relatively less accurate definition of H-bond), but the shape of the distribution, its bimodality, and the large separation between the two configurations are all well-conserved properties, highlighting the robustness of the structural description provided by $\Psi$.
The agreement between RD and IS apparently deteriorates on increasing $T$ (Fig.~\ref{fig:psird-isobar}). The differences arise from the mis-identification of H-bonds in the RD configurations at higher temperatures (needed to define the topological distances used to compute $\Psi$).
Indeed, as discussed in detail in the SI, at higher temperatures the distribution of molecular distances and orientations is widened by the increasing contribution of vibrational and librational modes, mixing the configurations corresponding to H-bonded and non-H-bonded pairs.
Above $\approx\SI{240}{\K}$, the two distributions (bonded and non-bonded, see Fig.~S3) are so widened by thermal motion that they superimpose in the region where the Luzar-Chandler cutoff acts. However, if the distribution of $\Psi$ is evaluated using the spatial configuration of the RD but retaining the H-bond network identified in the corresponding IS configuration (thick violet curves in Fig.~\ref{fig:psird-isobar}), then the same agreement between RD and IS that was observed at $T=\SI{188}{\K}$, is found at all temperatures.

At all the conditions we analyzed, the $\Psi^*$ distributions can be faithfully represented as a mixture of two Burr Type XII distributions~\cite{burr1942cumulative}
\begin{equation}
P(\Psi) = sP_L(\Psi;c_L, k_L, \lambda_L) + (1-s)P_H(\Psi;c_H, k_H, \lambda_H)
\label{eq:burr}
\end{equation}
whose parameters ($c, k,\lambda$) and relative weights ($s$) were optimized independently for each thermodynamic condition. These fits are superimposed as solid lines on the simulation data in Figures 1--4. The behavior of the distribution parameters (see SI) suggests that a non-trivial dependence on thermodynamic conditions is still present after removing the isotropic scaling component, preventing us from obtaining a simpler description of the behavior of $\Psi^*$.
 
\vskip 0.2cm

\begin{figure*}
	\includegraphics[width=.9\textwidth]{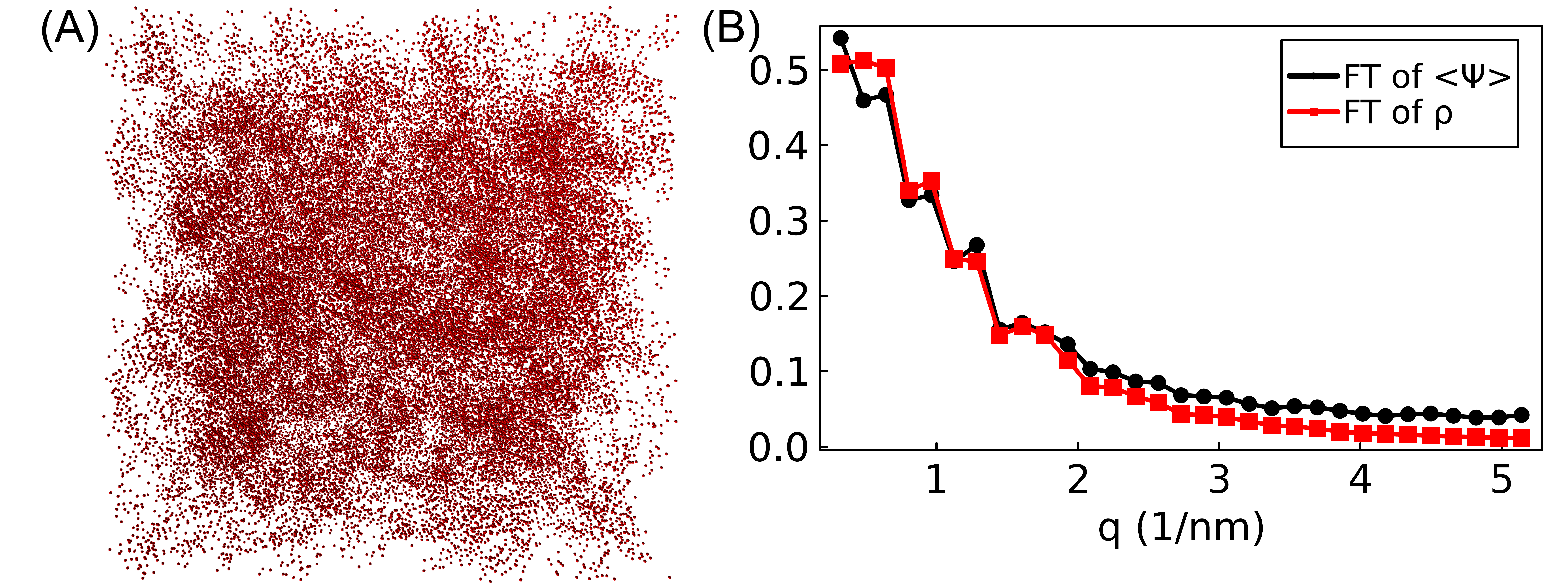}
	\caption{
        (A) Snapshot of molecules with low $\Psi^*$ from a constant-density simulation with 250000 molecules at $T=\SI{191}{\K}$ and $\rho=\SI{1.015}{\g\per\centi\m^3}$ visually shows spatial correlations
        extending over large length-scales.
        (B) Comparison between the Fourier transform of the averaged $\Psi^*$ field (multiplied along y by an arbitrary factor) and the
        Fourier transform of the density field $\rho$ (i.e., the structure factor), evaluated in the IS configurations, confirms that these two scalar
        fields show the same spatial correlations.
        The same calculation was also performed in the RD configurations (Fig.~S6) and shows no noticeable difference from the calculations in the IS.
    }
	\label{fig:CGcorrelations}
\end{figure*}

To provide a graphical representation of the spatial correlation of the 
molecules with similar $\Psi$ value, we perform and analyse a simulation
of a 250000 molecules systems, at $T=190$ K and $\rho=1.015$ g/cm$^3$, 
corresponding to the critical isochore of the TIP4P/Ice model. The 250000 molecule system are contained in  a cubic box of side of about 19.5 nm. Such a large distance, more than 60 times the nearest neighbour oxygen-oxygen distance (0.28 nm) makes it possible to
investigate the presence of long range correlations.  Fig.~\ref{fig:CGcorrelations}(a) shows with a
red sphere the position of the molecules with $\Psi^*<0.45$. Eyes immediately catch the
spatial correlation between them, with a correlation scale significantly larger than
the nearest neighbour distance. Such a correlation is a clear indication of a 
net attraction between molecules with similar environments. This observation can be
quantified and strengthened  by calculating the structure factor, the power spectrum of the  Fourier transform of the density, shown in Fig.~\ref{fig:CGcorrelations}B (evaluated in the IS configurations).
As expected close to the critical point, and as previously demonstrated for
the ST2 water model~\cite{guo2018anomalous} and also for TIP4P/2005 and TIP4P/Ice in smaller-sized systems~\cite{debenedetti2020second},
a strong increase in the scattered intensity at small $q$ is observed. 
To provide evidence that this correlation is brought up by the correlation between 
molecules with similar environment we define a $\expval{\Psi}$ field, by averaging
$\Psi_i$ over all molecules included in a small cubic volume. For convenience, we pick
this volume as $(L/32)^3$ where $L$ is the simulation box side. With this choice 
there are about 8 molecules in each mesh volume.  The resulting field $\expval{\Psi}$
can then be Fourier transformed in space and compared with $S(q)$.
The result of this comparison in shown in Fig.~\ref{fig:CGcorrelations}B, neatly demonstrating by the similarity of the two Fourier transforms that 
the density field  and the $\expval{\Psi}$ field not only provide the same information, but also that the spatial correlation in the density picked up by the small-angle scattering 
is identical to the spatial correlation of $\expval{\Psi}$.  In this respect, the two
families of molecules can be identified as the two states commonly assumed in 
mean field models.

\vskip 0.5cm
 

\section{Conclusions}

In this manuscript we have demonstrated that 
two distinct families of local environments can be clearly identified in numerical simulations of bulk liquid water on the basis of the structural indicator $\Psi_i$. 
This indicator  quantifies the
distance from a central molecule $i$ of the closest molecule separated by four hydrogen bonds.  In this respect, it requires information on the connectivity and local geometry  of the hydrogen bond network departing from each molecule.  The distribution of 
$\Psi_i$ becomes unimodal in two extreme cases, both at deep supercooling:
 at T=188 K, below the critical pressure 
 and at ambient pressure, all molecules belong to the large $\Psi$ family, while  at T=188 K and pressures above 10 kbar (when the configuration of water in the supercooled liquid is reminiscent of the very-high density amorphous structure~\cite{foffi2021structure}), all molecules belong to
the small $\Psi$ family.  Since the small $\Psi$ family is characterised by 
the presence of interstitial molecules between the first and the second tetrahedral shell, the very high density amorphous can be described as the limiting structure 
in which, despite the hydrogen bonds being mostly preserved (with only $\sim 10 \%$ of the molecules showing coordination defects at \SI{13}{\kilo\bar}~\cite{foffi2021structure}), all molecules are surrounded by and act as interstitial molecules.
The bimodal character of $\Psi$ is also conserved at ambient conditions, where, despite the increased thermal noise leading to a broadening of the two peaks and the existence of a single free energy minimum (at system level), it is still possible to clearly discern two structural families (at the single molecule level, in terms of $\Psi$).

The present findings provide a strong support to theoretical modelling of the 
thermodynamics of water as arising from the relative competition between these two families of local environments, each of them characterised by its own
local energy, density and entropy.  The water anomalies originate from the competition
between these two local structures, a phenomenon which is missing in simple liquids.

Finally, we have demonstrated that these local structures cluster in space. The analysis of a very large simulation, with more than 250000 molecules and providing access to distances extending up to 10 nm, allows us to demonstrate that
the correlation goes well beyond the 0.8-1 nm range which has long been associated to the typical decay of the spatial correlations in water~\cite{sciortino1989hydrogen}, thus confirming the preferential association of molecules of similar type, a feature which is essential for the existence of the liquid-liquid critical phenomenon.

\section*{Supplementary Material}
The supplementary material includes data on the phase diagram of the TIP4P/Ice model of water used in the present work, discussion on the identification of Hydrogen bonds, and additional data relative to the parameterization of the $\Psi^*$ distribution through Burr type XII distributions, and the comparison between inherent structures and real dynamics.

\begin{acknowledgments}
F.S. acknowledges support from 
PRIN 2022JWAF7Y,
ICSC—Centro Nazionale di Ricerca in High Performance Computing, Big Data and Quantum Computing, funded by the European Union—NextGenerationEU" and CINECA-ISCRAB.
\end{acknowledgments}

\section*{Data Availability Statement}

The data that support the findings of this study are available within the article. Additional data (MD trajectories) are available from the corresponding author upon reasonable request.

\bibliography{manuscript}

\end{document}



\setcounter{equation}{0}
\setcounter{figure}{0}
\setcounter{table}{0}
\setcounter{page}{1}
\makeatletter
\renewcommand{\theequation}{S\arabic{equation}}
\renewcommand{\thefigure}{S\arabic{figure}}
\renewcommand{\thetable}{S\arabic{table}}

\author{Riccardo Foffi}
\affiliation{Institute for Environmental Engineering, Department of Civil, Environmental and Geomatic Engineering, ETH Z\"urich, Laura-Hezner-Weg 7, 8093 Z\"urich, Switzerland.}
\author{Francesco Sciortino}
\affiliation{Dipartimento di Fisica, Sapienza Universit\`a di Roma, Piazzale Aldo Moro 5, I-00185 Rome, Italy.}

\title{Supplementary Information for: Identification of local structures in water from supercooled to ambient conditions}

\maketitle
\clearpage

\section*{Numerical simulations}
\begin{figure*}
	\includegraphics[width=10cm]{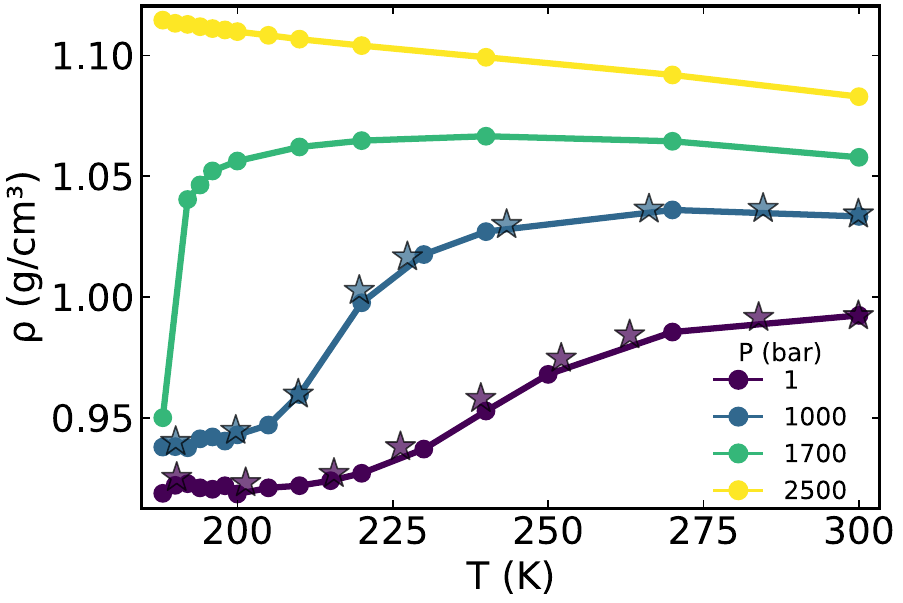}
	\caption{Equation of state $\rho(T)$ of TIP4P/Ice sampled by our simulations. The star markers are density estimates
    from TIP4P/Ice simulations at 1 and 1000 bar extracted from \citet{espinosa2023possible}.}
	\label{fig:EOS}
\end{figure*}

\begin{figure*}
	\includegraphics[width=10cm]{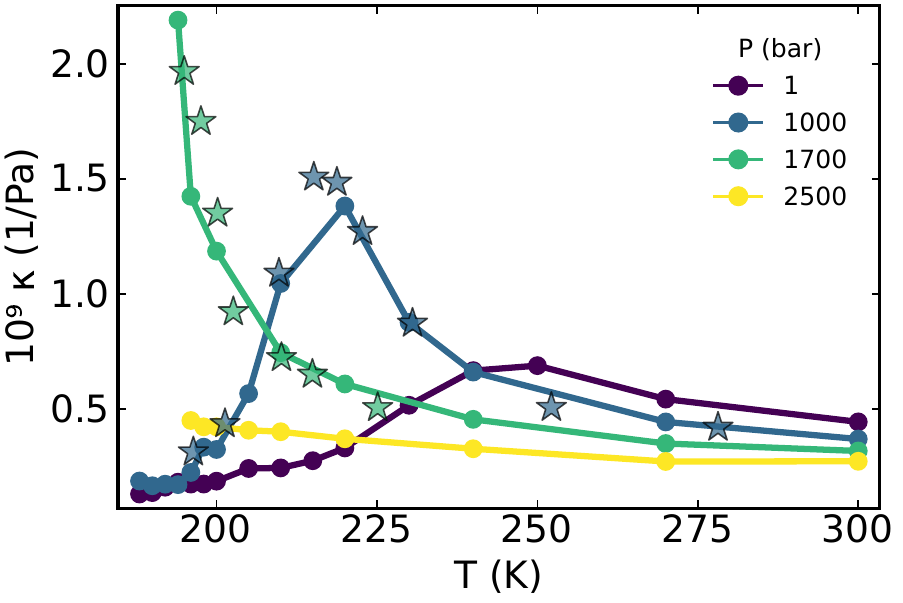}
	\caption{Isothermal compressibility of TIP4P/Ice sampled by our simulations. The star markers are compressibility estimates from TIP4P/Ice simulations at 1000 and 1700 bar extracted from \citet{espinosa2023possible}.}
	\label{fig:kTmax}
\end{figure*}

The results of our numerical simulations of TIP4P/Ice were successfully tested for reproducibility against the results reported in the recent work of \citet{espinosa2023possible}, as shown for the equation of state in Fig.~\ref{fig:EOS} and the isothermal compressibility in Fig.~\ref{fig:kTmax}, displaying the characteristic maxima.

\section*{Identification of Hydrogen bonds in the real dynamics}
\begin{figure*}
  \begin{center}
    \includegraphics[width=.9\textwidth]{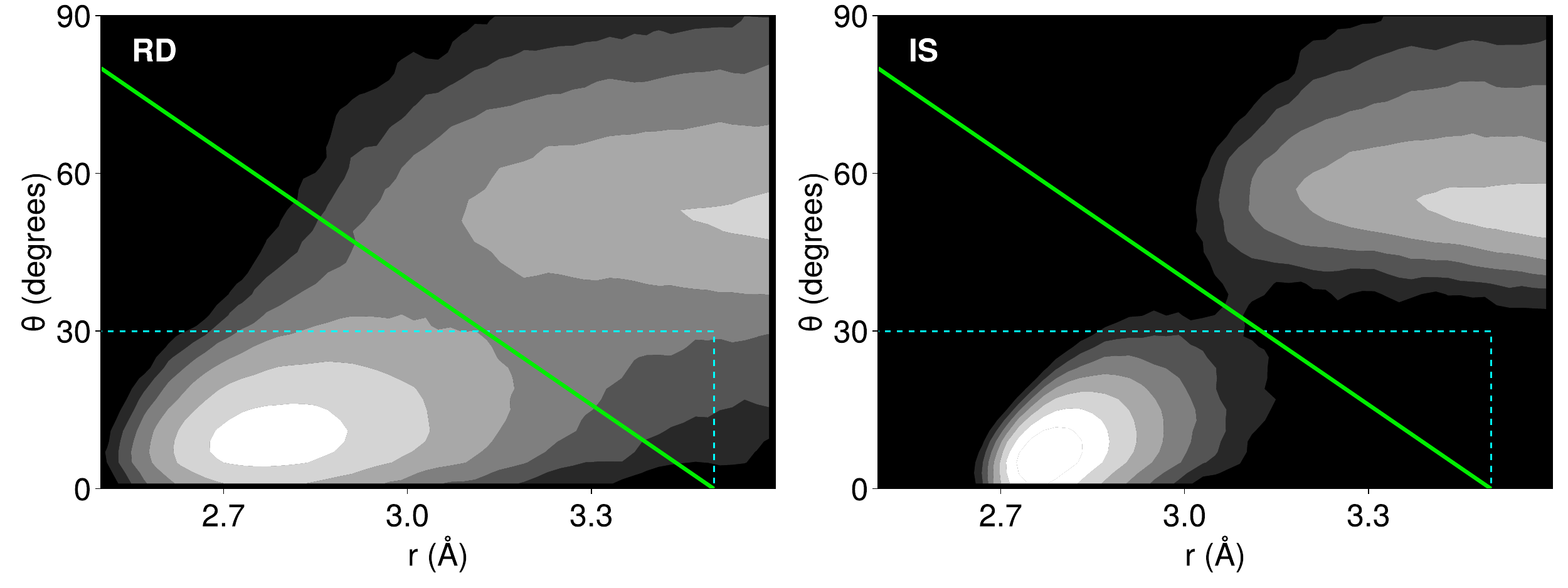}
  \end{center}
  \caption{Joint radial-angular pair distribution function of water molecules in (left) RD and (right) IS configurations at $T=\SI{300}{\K}$ and $P=\SI{1700}{\bar}$. The strong contribution of thermal motion at high temperatures complicates the definition of H-bonds in the RD.
  Dashed cyan lines represent the Luzar-Chandler definition of H-bond; solid green line is an ad-hoc definition obtained by defining a line of minimum cost passing through the saddle point of the pair distribution function.
  Level curves of the distributions are shown in logarithmic scale.}
  \label{fig:hbonds}
\end{figure*}

The vibrational and librational motion of water molecules, which is present in the RD but suppressed in the IS, significantly widens the radial-angular pair distribution function, adding a random thermal component that complicates the identification of H-bonds.
Fig.~\ref{fig:hbonds} shows the joint pair distribution of the Oxygen-Oxygen distance ($r$) and the intermolecular Hydrogen-Oxygen-Oxygen angle ($\theta$) for all pairs of molecules in the system.
In the IS, the identification of H-bonded pairs is obvious and unambiguous, as there is a well-separated basin which is perfectly captured by a simple definition such as that of Luzar and Chandler~\cite{luzar1993structure}: $r<\SI{3.5}{\angstrom}$ and $\theta<\SI{30}{\degree}$.
In the RD, while we can still identify these two basins, we are unable to clearly define a boundary between them to separate H-bonded and non-H-bonded pairs.
The two basins are now broadened by thermal motion and superimpose in the region where the Luzar-Chandler cutoff acts. Even using ad-hoc approaches, such as defining a diagonal cut through the saddle point of the pair distribution function, we are bound to mis-identify a non-negligible number of H-bonds, which will affect our definition of the H-bond network, and by extension the evaluation of $\Psi$.

\section*{Fitting the probability distribution of $\Psi^*$}
We tested a wide class of probability density functions to fit the behavior of $\Psi^*$ (Cauchy, Gamma, inverse Gamma, F, Frechet, Gumbel, Weibull, Burr and Skew-Normal distributions). The initial investigation was conducted by independently fitting the two limit distributions for the low-density ($T=\SI{188}{K}$, $P=\SI{0}{\bar}$) and the high-density ($T=\SI{188}{K}$, $P=\SI{13}{\kilo\bar}$) states for different distribution families.
For both states, the Burr Type XII distributions~\cite{burr1942cumulative} were found to provide the most accurate description of the data (although a good quality fit could also be obtained with Skew-Normal distributions):
\begin{equation}
    P(x) = \dfrac{ck}{\lambda}\qty(\dfrac{x}{\lambda})^{c-1}
    \qty[1 + \qty(\dfrac{x}{\lambda})^c]^{-k-1};
    \qquad x > 0,\, c > 0,\, k > 0,\, \lambda > 0.
\end{equation}

For each state point $(T,P)$, the distribution parameters were evaluated by optimizing over a binary mixture of the two distributions
\begin{equation}
    P(\Psi) = 
        sP_L(\Psi; c_L, k_L, \lambda_L) +
        (1-s)P_H(\Psi; c_H, k_H, \lambda_H)
\end{equation}

\begin{figure*}
    \centering
    \includegraphics[width=.9\textwidth]{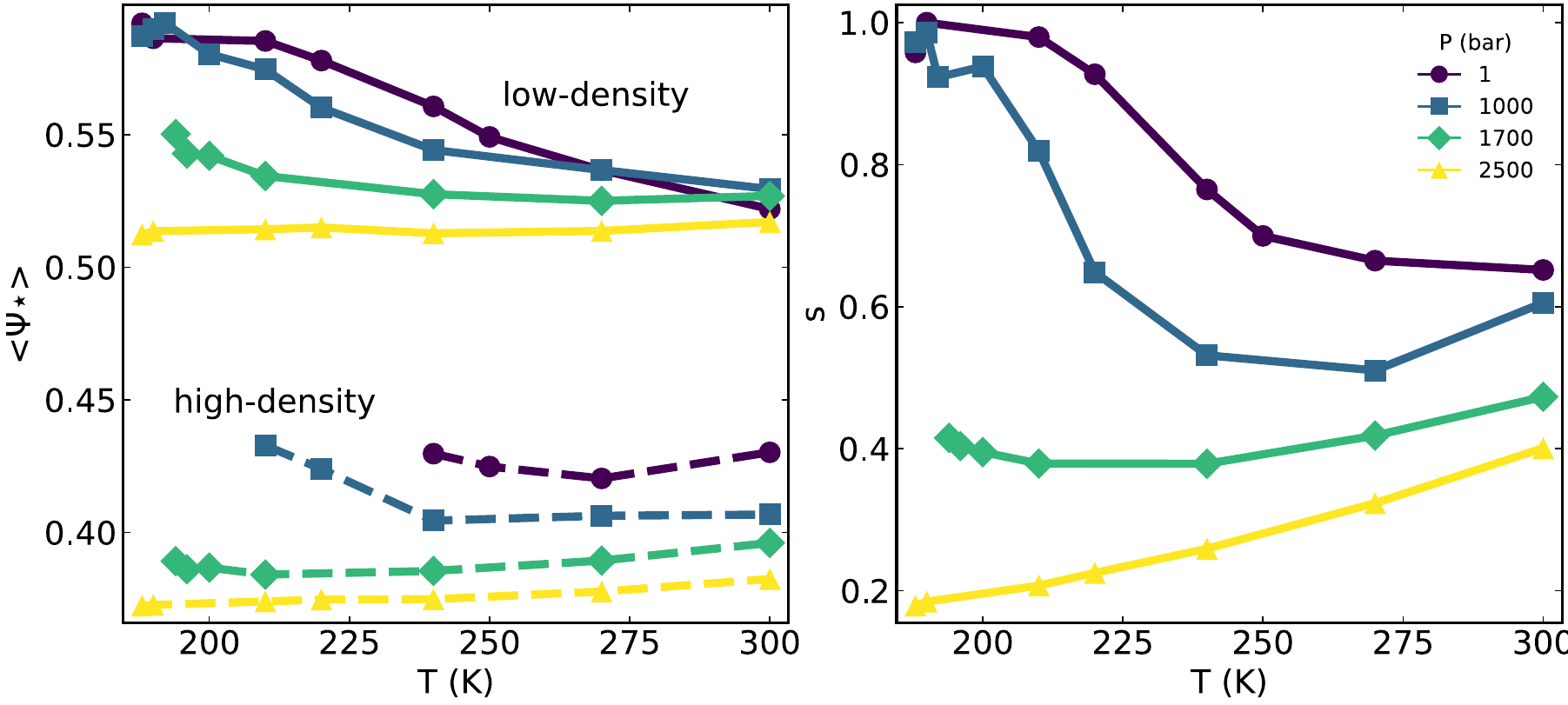}
    \caption{Temperature dependence of (A) the average of each structural family's distribution and (B) mixing parameter along the 4 isobars of Fig.~\ref{fig:EOS}, resulting from the fit of two Burr distributions.
    In panel A, the dashed lines represent the curves for the high-density component.
    At the lowest temperatures below the critical isobar, the fraction of high-density structure is extremely low ($\phi\simeq1$), so the corresponding parameter values are less significant:
    values of the HDL component are therefore not shown when $s>0.9$.}
    \label{fig:fitparameters}
\end{figure*}

The mixing parameter $s\in[0,1]$ represents the weight of the distribution associated to the LDL-like structure, it is therefore equivalent to what is generally identified as the order parameter or the fraction of locally-favored structures~\cite{russo2014understanding,tanaka2019revealing,foffi2023correlated}.
The other parameters which define the distributions, the two ``shape'' parameters $c$ and $k$ and the ``scale'' parameter $\lambda$, don't have an obvious physical meaning.
Therefore, the two quantities of interest with an immediate physical interpretations are the average values of each structural component's distribution and the mixing parameter, shown in Fig.~\ref{fig:fitparameters}.
For both high- and low-density structures, the distribution average decreases with increasing pressure, highlighting how pressure increases disorder within both types of structural arrangements, by reducing the average distance of molecules in the fourth coordination shell. The mixing parameter reflects the behavior displayed by the density in the equation of state (Fig.~\ref{fig:EOS}), confirming its accuracy as a possible order parameter.

\begin{figure*}
  \includegraphics[width=10cm]{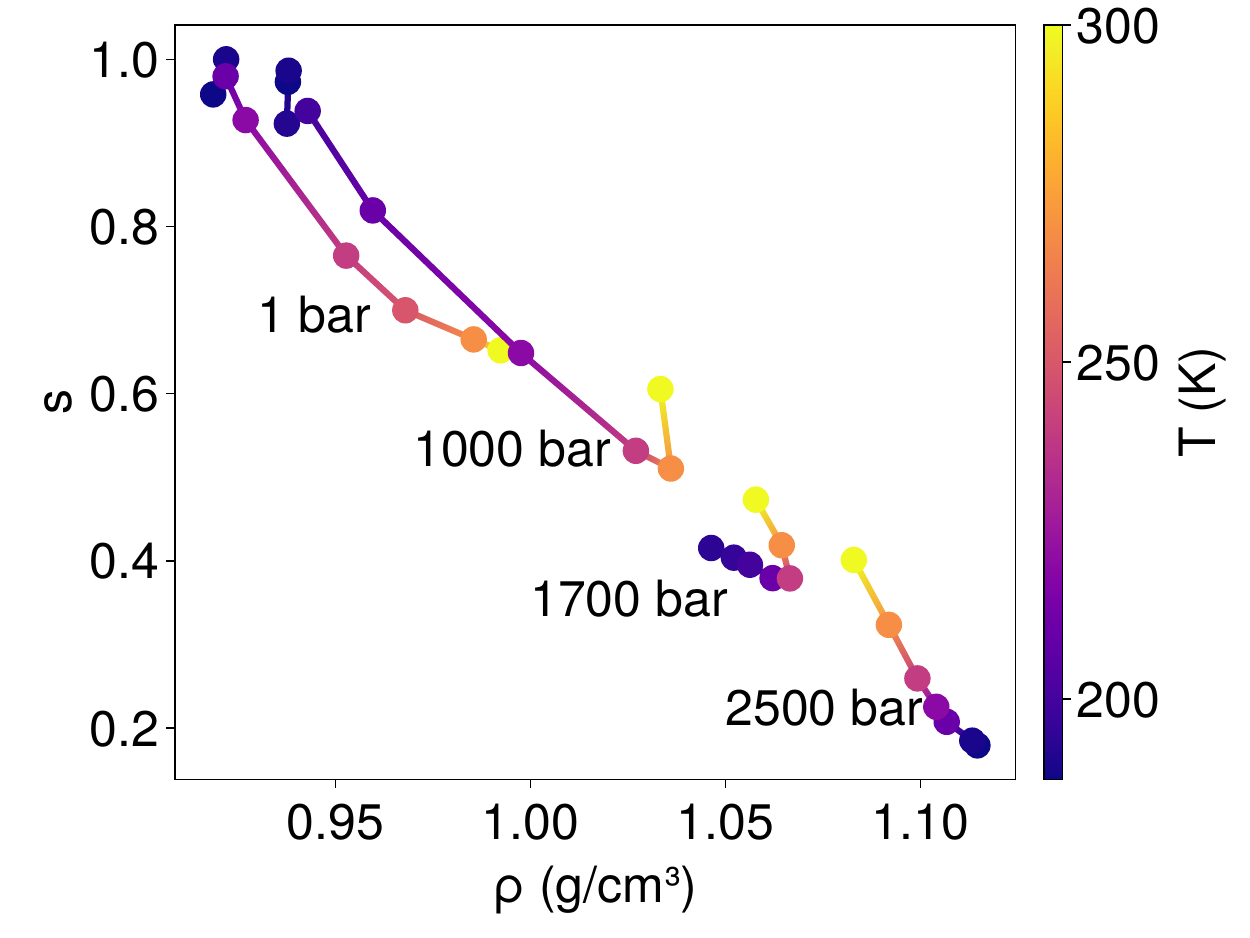}
  \caption{Fraction of locally favored structures $s$, evaluated by the regression of Burr Type XII distributions to the $\Psi^*$ distributions, shown as a function of the system density, $\rho$, for four different isobars (as indicated by annotations).
  The temperature along the isobars is color-coded following the colorbar on the right.}
  \label{fig:s-vs-rho}
\end{figure*}

The behavior displayed in Fig.~\ref{fig:fitparameters}B is indeed what would be expected for an order parameter which is tightly linked to the system density. When the isobars investigated in this work cross through temperature of maximum density (TMD) locus, then the inversion in the $\rho(T)$ dependency is reflected in $s(T)$. Fig.~\ref{fig:s-vs-rho} shows the same data for the fraction of locally favored structures, $s$, as a function of the system density $\rho$. At low $P$, an increase in $T$ leads to an increase in $\rho$, and therefore to a decrease in $s$. At intermediate $P$ the TMD line is crossed at $\approx\SI{270}{\K}$ and $\approx\SI{240}{\K}$ (for \SI{1000}{\bar} and \SI{1700}{\bar} respectively), showing that minimum $s$ corresponds to maximum $\rho$. Finally, the $P=\SI{2500}{\bar}$ isobar lies entirely above the TMD line, so that with increasing $T$, $\rho$ decreases, and $s$ increases.

We note that the position of the inflection point in the $T$-dependence of the mixing parameter $s$ along the \SI{1}{\bar} and (to a lesser extent) the \SI{1000}{\bar} isobars (see Fig.~\ref{fig:fitparameters}B) does not coincide with the temperatures at which the isothermal compressibility has a maximum. In our case, the value of $s$ at the compressibility maxima is larger than $1/2$. In terms of two-state models, this could indicate either the presence of clustering in the two molecular local arrangements or a non-negligible contribution to the compressibility of the reference state, more relevant when far from the critical point. A future analysis of TIP4P/Ice with a two-state model which incorporates correlations between the local environments~\cite{holten2013nature,daidone2023statistical} may help shed light on these observations.

The parameters resulting from our fits, for all the thermodynamic conditions analyzed, are reported in Tables S1 and S2.

\section*{Long-range correlations in the real dynamics}
\begin{figure*}
  \includegraphics[width=10cm]{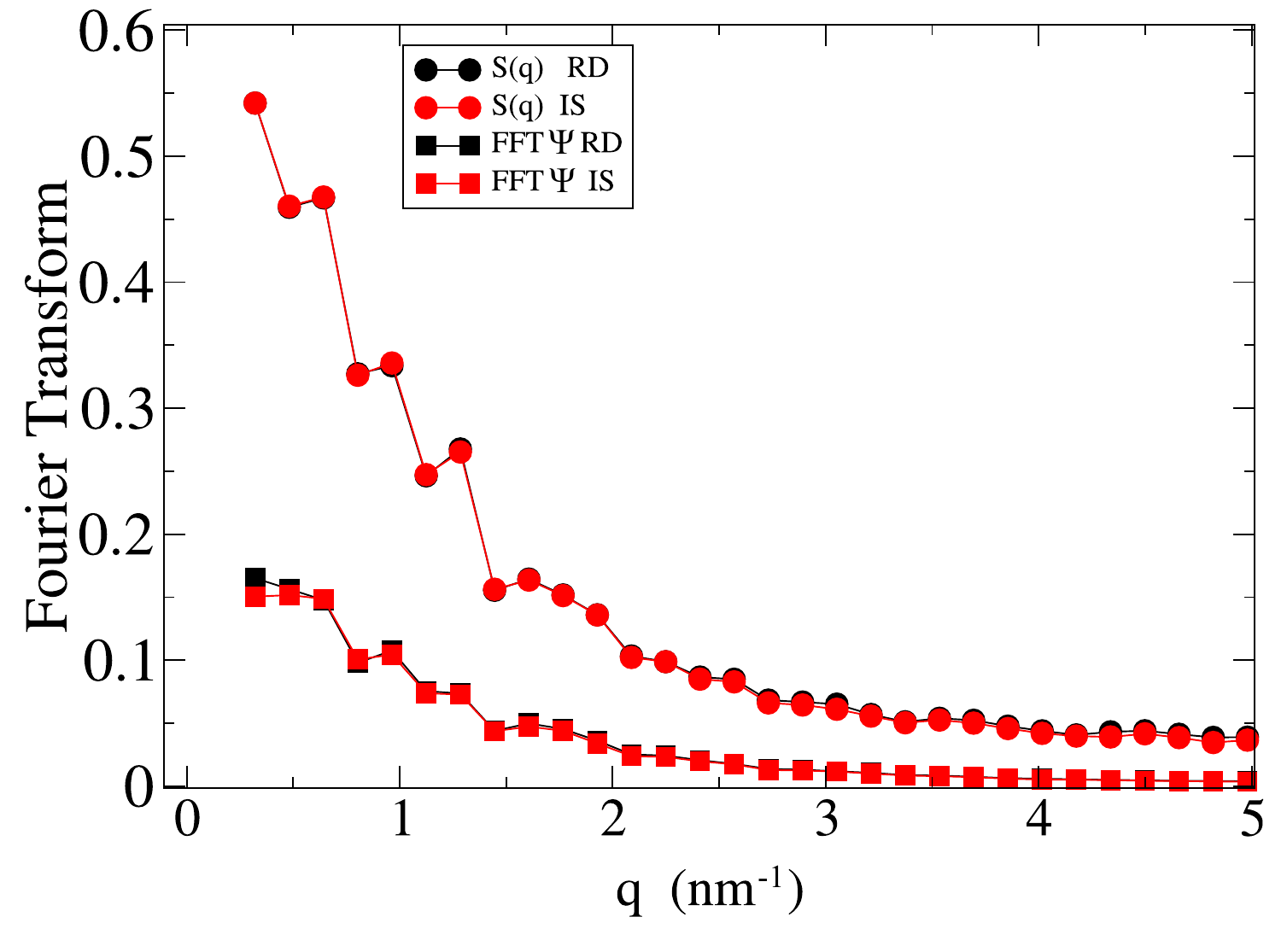}
  \caption{Comparison of the Fourier transforms of the density field (structure factor) and of the coarse-grained $\expval{\Psi^*}$ field between the RD and IS configurations, evaluated for a system of $N=250000$ molecules at $T=\SI{190}{\K}$ and $\rho=\SI{1.015}{\g\per\centi\meter^3}$.}
  \label{fig:sq}
\end{figure*}

Fig.~\ref{fig:sq} shows the Fourier transform of the coarse-grained $\expval{\Psi^*}$ field and of the density field (i.e., the structure factor) evaluated in the RD, comparing them to their values in the IS as shown in Fig.6B in the main text. Indeed, the energy minimization procedure only produces local changes in the coordinates of the molecules, leaving the low-$q$ density fluctuations unaffected. Similarly, close to the critical point, hydrogen bonds are properly identified both in the RD and the IS, yielding identical estimates of $\Psi$.

\clearpage
\begin{table}
\begin{tabularx}{\textwidth}{*{9}{X}}
\toprule[2pt]
$T$ (K) & $P$ (bar) & $c_L$ & $k_L$ & $\lambda_L$ & $c_H$ & $k_H$ & $\lambda_H$ & $s$ \\
\midrule[1pt]
188.0 & 0.0 & 16.25 & 5.949 & 0.6752 & 16.16 & 0.598 & 0.4602 & 0.958 \\
188.0 & 1000.0 & 14.31 & 36.66 & 0.779 & 11.95 & 0.591 & 0.4079 & 0.973 \\
188.0 & 11000.0 & 12.88 & 42.13 & 0.7468 & 60.6 & 0.2565 & 0.3391 & 0.07635 \\
188.0 & 13000.0 & 3.208 & 20.92 & 1.514 & 60.97 & 0.2925 & 0.3398 & 0.09002 \\
188.0 & 1675.0 & 12.31 & 81.33 & 0.8625 & 15.81 & 0.3064 & 0.3572 & 0.8379 \\
188.0 & 1775.0 & 14.07 & 80.08 & 0.836 & 8.024 & 1.782 & 0.471 & 0.8851 \\
188.0 & 1776.0 & 11.67 & 186.3 & 0.9189 & 37.15 & 0.1095 & 0.3319 & 0.3742 \\
188.0 & 1801.0 & 11.95 & 124.0 & 0.8778 & 36.66 & 0.1119 & 0.3329 & 0.3473 \\
188.0 & 1901.0 & 12.36 & 119.5 & 0.861 & 39.16 & 0.1112 & 0.333 & 0.2932 \\
188.0 & 2500.0 & 14.07 & 127.3 & 0.8207 & 50.92 & 0.1009 & 0.3315 & 0.1791 \\
188.0 & 5000.0 & 12.93 & 78.42 & 0.8011 & 56.22 & 0.1657 & 0.3347 & 0.1261 \\
188.0 & 7000.0 & 12.54 & 27.41 & 0.7389 & 57.56 & 0.2087 & 0.3368 & 0.1042 \\
188.0 & 9000.0 & 11.79 & 86.68 & 0.8212 & 61.36 & 0.2264 & 0.3383 & 0.08738 \\
190.0 & 0.0 & 13.47 & 52.14 & 0.8149 & 61.70 & 0.5982 & 0.3991 & 1.0 \\
190.0 & 1000.0 & 13.29 & 71.81 & 0.8426 & 20.06 & 0.5614 & 0.3852 & 0.9863 \\
190.0 & 1500.0 & 12.69 & 83.11 & 0.859 & 9.253 & 1.06 & 0.4262 & 0.8892 \\
190.0 & 1650.0 & 12.93 & 69.47 & 0.8415 & 9.565 & 0.7417 & 0.4054 & 0.8215 \\
190.0 & 1725.0 & 11.76 & 133.9 & 0.9094 & 26.48 & 0.1433 & 0.3326 & 0.6441 \\
190.0 & 1800.0 & 11.66 & 100.7 & 0.8751 & 37.3 & 0.1058 & 0.3313 & 0.39 \\
190.0 & 1900.0 & 12.38 & 126.2 & 0.8629 & 41.55 & 0.1024 & 0.3319 & 0.2843 \\
190.0 & 2000.0 & 12.34 & 136.0 & 0.8676 & 40.9 & 0.1088 & 0.3321 & 0.2698 \\
190.0 & 2500.0 & 13.94 & 81.55 & 0.797 & 48.58 & 0.1065 & 0.3319 & 0.1847 \\
\end{tabularx}
\caption{Parameters resulting from the fit of the $\Psi^*$ distributions with a binary
mixture of two Burr Type XII distributions (part 1).}
\end{table}

\begin{table}
\begin{tabularx}{\textwidth}{*{9}{X}}
\toprule[2pt]
$T$ (K) & $P$ (bar) & $c_L$ & $k_L$ & $\lambda_L$ & $c_H$ & $k_H$ & $\lambda_H$ & $s$ \\
\midrule[1pt]
192.0 & 1000.0 & 14.61 & 15.67 & 0.7339 & 10.59 & 4.88 & 0.5564 & 0.923 \\
194.0 & 1700.0 & 11.27 & 137.4 & 0.9078 & 32.71 & 0.1246 & 0.3336 & 0.4153 \\
196.0 & 1700.0 & 10.87 & 177.1 & 0.9375 & 34.94 & 0.1185 & 0.3327 & 0.4037 \\
200.0 & 1000.0 & 13.19 & 59.69 & 0.8239 & 7.874 & 2.259 & 0.4872 & 0.9382 \\
200.0 & 1700.0 & 10.77 & 207.2 & 0.9559 & 36.17 & 0.1119 & 0.3321 & 0.3952 \\
210.0 & 0.0 & 13.72 & 59.89 & 0.8152 & 10.94 & 2.954 & 0.4657 & 0.9797 \\
210.0 & 1000.0 & 12.46 & 46.64 & 0.8169 & 9.457 & 0.8321 & 0.4122 & 0.8192 \\
210.0 & 1700.0 & 10.54 & 161.9 & 0.9346 & 37.49 & 0.1102 & 0.331 & 0.3792 \\
210.0 & 2500.0 & 12.76 & 205.5 & 0.8764 & 48.11 & 0.09701 & 0.3299 & 0.2073 \\
220.0 & 0.0 & 13.14 & 86.5 & 0.8455 & 7.929 & 2.072 & 0.4986 & 0.9273 \\
220.0 & 1000.0 & 10.55 & 138.5 & 0.9356 & 18.12 & 0.2457 & 0.3457 & 0.6486 \\
220.0 & 2500.0 & 11.66 & 135.9 & 0.8722 & 50.78 & 0.08798 & 0.3286 & 0.2252 \\
240.0 & 0.0 & 11.76 & 52.34 & 0.8266 & 9.371 & 0.9294 & 0.42 & 0.7651 \\
240.0 & 1000.0 & 9.269 & 85.13 & 0.9204 & 29.28 & 0.1411 & 0.3323 & 0.5316 \\
240.0 & 1700.0 & 9.384 & 197.3 & 0.9913 & 35.76 & 0.1211 & 0.3304 & 0.3789 \\
240.0 & 2500.0 & 10.38 & 159.9 & 0.9214 & 45.22 & 0.1018 & 0.3287 & 0.2592 \\
250.0 & 0.0 & 10.85 & 135.0 & 0.9166 & 11.07 & 0.6345 & 0.3893 & 0.6997 \\
270.0 & 0.0 & 9.042 & 168.3 & 0.9983 & 20.45 & 0.2244 & 0.3398 & 0.6648 \\
270.0 & 1000.0 & 8.493 & 155.6 & 1.01 & 28.75 & 0.1484 & 0.3313 & 0.5104 \\
270.0 & 1700.0 & 8.299 & 177.9 & 1.032 & 38.76 & 0.1118 & 0.3284 & 0.4185 \\
270.0 & 2500.0 & 8.988 & 95.34 & 0.9214 & 43.62 & 0.1044 & 0.3279 & 0.3233 \\
300.0 & 0.0 & 8.286 & 103.7 & 0.9693 & 19.24 & 0.2616 & 0.3443 & 0.6518 \\
300.0 & 1000.0 & 7.88 & 11.13 & 0.7365 & 24.07 & 0.2425 & 0.3356 & 0.6054 \\
300.0 & 1700.0 & 7.856 & 97.14 & 0.9746 & 34.8 & 0.1302 & 0.3292 & 0.4731 \\
300.0 & 2500.0 & 8.048 & 100.3 & 0.9667 & 40.76 & 0.1131 & 0.328 & 0.401 \\
\end{tabularx}
\caption{Parameters resulting from the fit of the $\Psi^*$ distributions with a binary
mixture of two Burr Type XII distributions (part 2).}
\end{table}

\clearpage
\bibliography{si}